\documentclass[aps,twocolumn,showpacs,amsmath,amsfonts,amssymb]{revtex4}
\usepackage{graphicx}
\usepackage{epsf}

\begin{document}

\title{Sub-Planck scale structures in the
P{\"o}schl-Teller potential and their sensitivity to
perturbations}

\author{Utpal Roy,$^{\mathrm{1}}$\footnote{e-mail:
utpal.roy@unicam.it} Suranjana
Ghosh,$^{\mathrm{1}}$\footnote{e-mail: suranjana.ghosh@unicam.it}
P. K. Panigrahi,$^{\mathrm{2,3}}$\footnote{e-mail:
prasanta@prl.res.in} and David
Vitali,$^{\mathrm{1}}$\footnote{e-mail: david.vitali@unicam.it}}
\affiliation{$^{\mathrm{1}}$Dipartimento di Fisica, Universit\`{a}
di Camerino, I-62032 Camerino, Italy\\$^{\mathrm{2}}$Indian
Institute of Science Education and Research Kolkata,
India\\$^{\mathrm{3}}$Physical Research Laboratory, Navarangpura,
Ahmedabad, India}

\begin{abstract}
We find the existence of sub-Planck scale structures in the
P{\"o}schl-Teller potential, which is an exactly solvable
potential with both symmetric and asymmetric features. We analyze
these structures in both cases by looking at the Wigner
distribution of the state evolved from an initial coherent state
up to various fractional revival times. We also investigate the
sensitivity to perturbations of the P{\"o}schl-Teller potential
and we verify that, similar to the harmonic oscillator, the
presence of sub-Planck structure in phase space is responsible for
a high sensitivity to phase-space displacements.
\end{abstract}

\pacs{03.65.-w, 42.50.Dv, 42.50.Md} \maketitle

\section{Introduction}

The Wigner distribution $W(x,p)$ is a useful tool for visualizing
quantum interference phenomena in phase space. Non-local
superposition states show interference and the presence of phase
space regions where $W(x,p)$ is negative is an unambiguous
signature of quantum behavior. The oscillatory structures
resulting from quantum interference can have a dimension much
smaller than the Planck's constant $\hbar$, and these sub-Planck
scale structures were first found by Zurek in quantum chaotic
systems \cite{zurek}. Sub-Planck scale structures usually appear
as alternate small 'tiles' of maxima and minima. The phase space
area of these 'tiles' are given by $\hbar^2$ divided by the
effective phase-space area $A$ globally occupied by the state,
which can be significantly larger than $\hbar$. These structures
are very sensitive to environmental decoherence
\cite{zurek,ph,wisn,pathak1,pathak2,ghosh1} and may have important
application in Heisenberg-limited measurements and quantum
parameter estimation \cite{toscano,dalvit}. A classical analogue
of sub-Planck scale structures, regarded as sub-Fourier
sensitivity, has been studied in Ref.~\cite{praxmeyer}, and these
structures have also been found in diatomic molecular systems
\cite{ghosh}, entangled cat states \cite{manan}, optical fibers
\cite{fibre}, and Kirkwood-Rihaczek distribution \cite{jay}.
Recently, a connection between these structures and teleportation
fidelity has been also established \cite{scott}.

Sub-Planck structures have been extensively studied in the case of superpositions of harmonic oscillator coherent states and also for
superpositions of generalized coherent states which occur in the time evolution of systems with nonlinear potentials at fractional revival
times. It is therefore important to extend the study of this phenomenon to other nonlinear potentials in order to investigate which are the
relevant parameters affecting this subtle quantum behavior, and if Heisenberg-limited sensitivity can be approached also in other nonlinear
systems.

Here we focus our attention to the P\"{o}schl-Teller (PT) potential, which has been applied to model various situations in recent years
\cite{tong1,tong2,indjin,tomak1,tomak2}. In fact, the PT potential has been used in the context of deep quantum wells, as well as for modeling
optical systems with changing refractive index. The PT potential depends upon two parameters and by varying them one can switch from a symmetric
to an asymmetric situation. Yildirim and Tomak studied in particular the nonlinear optical properties associated to the PT potential
\cite{tomak1}, and also the nonlinear changes in the refractive index resulting from its tunable asymmetry \cite{tomak2}.

Another important feature of the PT potential is that it is exactly solvable and characterized by a quadratic spectrum: as a consequence its
time evolution shows fractional revivals \cite{averbukh,robinett} and it may lead to the generation of linear superpositions of well distinct
states in phase space. In this paper we confirm this expectation and find that an initial coherent state of the PT potential shows sub-Planck
structures in phase-space at fractional revival times. We shall first study these structures when the PT potential is symmetric and then, by
tuning the potential parameters, we switch to the asymmetric case in order to see the effects of the asymmetry on sub-Planck scale structures.
In both cases we verify quantitatively that the dimension of the sub-Planck structures scales as $\hbar^2/A$, as expected, also for the PT
potential.

We then study the sensitivity of sub-Planck scale structures to perturbations, by considering two different situations: i) ``internal''
perturbations, i.e., a small change of the potential which is turned from symmetric into a slightly asymmetric one; ii) ``external''
perturbations, as for example a weak force, yielding a small phase space displacement of the state. Both cases will confirm that also in the PT
potential, sub-Planck structures are responsible for a very high sensitivity to displacements.

The paper is organized as follows. In Sec II, the symmetric and asymmetric properties of PT potential are reviewed and the revival dynamics of a
coherent state (CS) wave-packet are discussed. Existence of sub-Planck scale structures in phase space is demonstrated for both symmetric and
asymmetric cases in Sec III. In Sec IV we discuss the sensitivity of sub-Planck scale structures due to a small asymmetry of the potential and
then the one associated with a small phase-space displacement of the state. We end with some conclusions in Sec V.

\begin{figure}[htpb]
\centering
\includegraphics[width=3.2in]{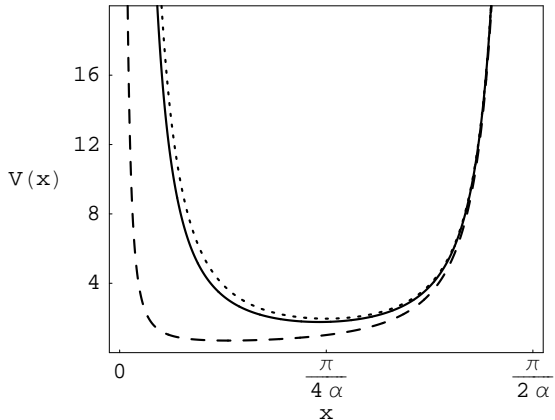}
\caption{The P\"{o}schl-Teller potential V(x) (in unit of $10^4\;a.u.$) for different parameter values. Dotted line represents the symmetric
case ($\rho=\kappa=50$), solid line refers to the slightly asymmetric case ($\rho=50$, $\kappa=46$), while the strongly asymmetric case
($\rho=50$ $\kappa=6$) is shown by the dashed line. Here, the variable $x$ is in atomic units and we have chosen $\alpha=2\;a.u$.
}\label{potential}
\end{figure}

\section{P\"{o}schl-Teller potential and the CS wave-packet}

The general PT potential \cite{posh} is given by
\begin{equation}
V(x)\;=\;\frac{\hbar^2\alpha^2}{2m}\left[\frac{\rho(\rho-1)}{\cos^2\alpha x}+\frac{\kappa(\kappa-1)}{\sin^2\alpha x} \right] \label{PT}
\end{equation}
and it represents an array of potential wells in the whole line.
The symmetric case can exhibit energy bands \cite{ranjani}. Since
the wells are impenetrable, it suffices to consider only one of
them for a full quantum mechanical description of the system. The
potential is depicted in Fig.~\ref{potential} for different values
of $\rho$ and $\kappa$ $(\rho,\kappa>1)$ (units are chosen such
that $\hbar=m=1$), and we have selected the well within the
interval $[0,\pi/2\alpha]$. The dotted line corresponds to the
symmetric well $(\rho=\kappa=50)$, centered around $\pi/4\alpha$.
This symmetric well smoothly takes the form of an infinite well in
the limit of $\rho,\kappa\rightarrow 1.$ The potential loses its
symmetry with respect to the axis of the well if $\rho \neq
\kappa$. Small and large asymmetries are shown respectively by the
solid and dashed curves in Fig.~\ref{potential}. The minimum of
the well is inclined towards the left side when $\rho>\kappa$.

The PT potential is one of the exactly solvable quantum mechanical potentials. The corresponding Schr\"{o}dinger equation can be solved to
obtain the energy eigenvalues and eigenfunctions \cite{klauder}
\begin{eqnarray}
E_{n}&=&\frac{\hbar^2\alpha^2}{2m} (2n+\rho+\kappa)^2,\quad
n=0,1,2,... ~~\nonumber\\
\psi_{n}(x)&=&N
\cos^{\rho}(\alpha x) \sin^{\kappa}(\alpha x) P_n^{\rho-1/2,\kappa-1/2} [\cos(2\alpha x)],\nonumber\\
\label{PTeigenfunction}
\end{eqnarray}
where $P_n^{\rho-1/2,\kappa-1/2} [\cos(2\alpha x)]$ is the Jacobi polynomial and $N$ is the normalization constant:
\begin{equation}
N=\left[\frac{2 \alpha
\Gamma{(n+1)}(2n+\rho+\kappa)\Gamma{(n+\kappa+\rho)}}{\Gamma{(n+\rho+1/2)}
\Gamma{(n+\kappa+1/2)}}\right]^\frac{1}{2}.
\end{equation}
It is worth mentioning that the PT potential has an underlying $SU(1,1)$ dynamical symmetry algebra, which admits an infinite number of energy
levels. Some of them can be excited to create a CS wave-packet by using an appropriate laser. These states have many classical features and tend
to preserve their properties even for long evolution times. Here, we consider a displacement operator CS \cite{perelomov} of this potential.
Construction of this CS makes use of the exponential form for the solution of hypergeometric differential equation and the correct choice of
$SU(1,1)$ generators \cite{guru1}. The CS depends upon a parameter $\beta$ and can be written as \cite{utpal}
\begin{equation}
\chi_{\beta}(x)\;=\;\sum_{n=0}^{\infty}\;\;d_{n}^{\beta}\;\psi_{n}(x),
\end{equation}
where the amplitudes $d_{n}^{\beta}$ are given by
\begin{equation}
d_{n}^{\beta}\;=\;(-\beta)^n
\left[\frac{\Gamma{(\rho+n+1/2)}\Gamma{(\kappa+n+1/2)}}{2 \alpha
\Gamma{(\kappa+\rho+n)}\Gamma{(n+1)}(2n+\kappa+\rho)}\right]^{1/2}.
\end{equation}
The eigenenergy $E_{n}$ has both linear and quadratic terms in the quantum number $n$. The time evolution of the CS generated by the PT
Hamiltonian,
\begin{equation}
\chi_{\beta}(x,t)=\sum_{n=0}^{\infty}d_{n}^{\beta}\psi_{n}(x)e^{-iE_nt}, \label{timeevolution}
\end{equation}
is such that, at short times, the center of the wave-packet
reproduces the classical motion, due to the linear term in $n$ in
the energy eigenvalues. The quadratic term in $n$ is instead
responsible for revival and fractional revivals
\cite{averbukh,robinett}, taking place on a longer time scale. In
fact, it is easy to check that after the revival time
$T_{rev}=\pi/\alpha^2$ (or integer multiples of it), the initial
coherent state localizes back to its form after spreading. Here
however, we are interested in fractional revivals, which instead
take place after a time interval $(r/s)T_{rev}$, with $r$ and $s$
mutually prime integers. At fractional revival times the initial
wave-packet partially localizes into a superposition of spatially
distributed sub-packets, each of which closely resembles the
initial wave-packet: when $s$ is even, the wave-packet breaks into
$s/2$ wave-packets, when instead $s$ is odd, the state is a
superposition of $s$ distinct states. This means that one obtains
a Schr\"{o}dinger cat state, superposition of two CS with opposite
phases, at one fourth of the revival time, while four-way break
up, or the so called ``compass-like'' state emerges at
$T_{rev}/8$. As shown by Ref.~\cite{zurek}, the latter produces
sub-Planck scale structures in the phase space Wigner
distribution. The behavior at fractional revival times can be
derived in a straightforward way from Eq.~(\ref{timeevolution}) in
the particular case when both $\rho$ and $\kappa$ are even. When
the PT potential is symmetric ($\rho=\kappa$), the property of the
Jacobi polynomial allows us to write
\begin{eqnarray}\label{symm}
&&\psi_{n}(x)=(-1)^{n}\psi_{n}(\pi/2\alpha-x)\nonumber\\
\quad &or&\;\;
\psi_{n}(\pi/4\alpha-x)=(-1)^{n}\psi_{n}(\pi/4\alpha+x).
\end{eqnarray}
Using this fact at $T_{rev}/4$, a straightforward
calculation yields
\begin{equation}
\chi_{\beta}(x,T_{rev}/4)=\frac{1}{\sqrt{2}}\left[e^{-i\pi/4}\chi_{\beta}(x,0) + e^{i\pi/4}\chi_{\beta}(\frac{\pi}{2\alpha}-x,0)\right],
\end{equation}
that is, the CS is split into two parts at time $T_{rev}/4$
forming the well known Schr\"{o}dinger cat state. Each element of
the superposition is proportional to the initial wave-packet, even
though they are situated at the opposite ends of the potential
well. In the asymmetric case instead one gets
\begin{widetext}
\begin{eqnarray}
&&\chi_{\beta}(x,T_{rev}/4)=\chi_{\beta}^{e}(x,0)- i\chi_{\beta}^{o}(x,0);\;\;\quad  \eta=\rho+\kappa=\mathrm{multiple\;of}\;
4\nonumber\\
\mathrm{and}\;\;&&\chi_{\beta}(x,T_{rev}/4)=-i\chi_{\beta}^{e}(x,0)+
\chi_{\beta}^{o}(x,0);\;\;\quad\eta=\mathrm{multiple\;of}\; 2,
\end{eqnarray}
\end{widetext}
where $\chi_{\beta}^{e,o}(x,0)=\sum_{n,\; even,odd} d_n \psi_n
(x)$. For the analysis at time $T_{rev}/8$, it is instead
convenient to define a classical wave-packet
\cite{averbukh,robinett}:
\begin{equation}
\chi_{\beta}^{cl}(x,t)=\sum_{n=0}^{\infty}d_{n}^{\beta}\psi_{n}(x)e^{-2\pi
i n t/T_{cl}}, \label{cl}
\end{equation}
which evolves with the classical periodicity $T_{cl}=\pi/(\rho+\kappa)\alpha^{2}$. It can be easily seen that at $t=T_{rev}/2$, the CS will
revive with an extra half a period ($T_{cl}/2$) of phase. In other words, the CS will be situated on the other side of the potential well, but
with the same amplitude. The wave-packet splits into two parts at time $t=T_{rev}/4$ with a relative phase of half a period $T_{cl}/2$. Four way
break-up occurs instead at time $t=T_{rev}/8$, and one can see that the state at this fractional revival time becomes
\begin{widetext}
\begin{equation}\label{oneeightcl}
\chi_{\beta}(x,T_{rev}/8)=\frac{1}{2}\left[e^{-i\pi/4}\chi_{\beta}^{cl}(x,t)
+\chi_{\beta}^{cl}(x,t+T_{cl}/4)-e^{-i\pi/4}\chi_{\beta}^{cl}(x,t+T_{cl}/2)+
\chi_{\beta}^{cl}(x,t+3T_{cl}/4)\right].
\end{equation}
\end{widetext}
One can generalize in a straightforward way to a generic fractional revival time. In fact, in this case the state can be written as
 \begin{equation}
\chi_{\beta}(x,t=\frac{r}{s}T_{rev})=\sum_{p=0}^{l-1}a_{p}\chi_{\beta}^{cl}(x,t-\frac{p}{l}T_{cl}),
\end{equation}
where $a_{p}$ gives the amplitude with probability $|a_{p}|^{2}$ of each $l$ number of sub-wave-packets or clones of the original wave-packet.
Each clone differs in phase from the original one by $T_{cl}/l$. As illustrated above, the six way break-up can be observed at the time
$T_{rev}/12$.

\begin{figure*}[htbp] \centering
  \includegraphics[width=7.in]{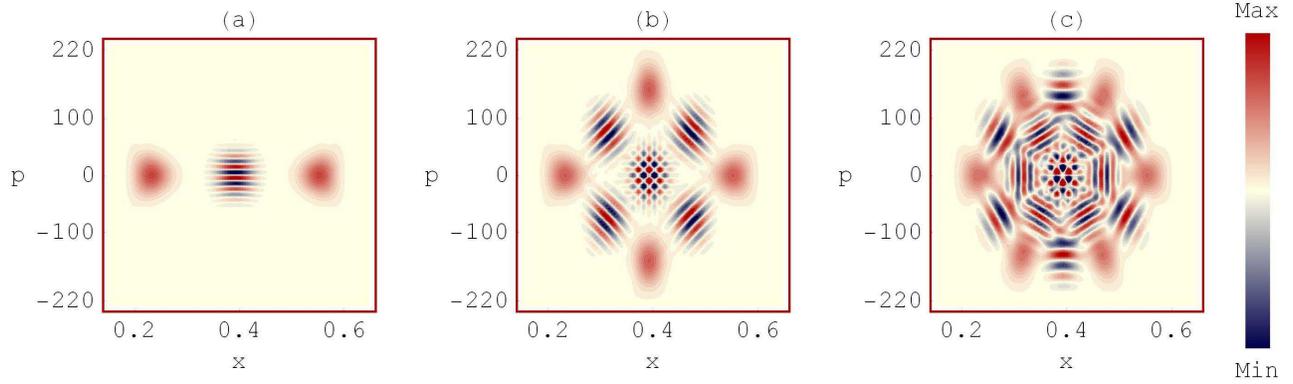}
\caption{(Color online) Contour plots of the Wigner distribution
of the time evolution of the CS, at different fractional revival
times, for a symmetric PT potential. (a) cat-like state at
$T_{rev}/4$; (b) compass-like state at $T_{rev}/8$; (c)
benzene-like state at $T_{rev}/12$. Here we have chosen
$\rho=\kappa=50$, $\alpha=2\;a.u.$, $\beta=0.6$ and $x$, $p$ are
the conjugate variables in atomic units.} \label{wigSymAll}
\end{figure*}

\section{Wigner distribution and sub-Planck scale structure}

The Wigner quasi-probability distribution provides a clear phase-space description of a quantum state~\cite{schleich}. It is particularly suited
for displaying quantum interference phenomena, which are associated to oscillations and to the presence of regions where the Wigner function
attains negative values. From its definition, the Wigner function of the CS evolved up to time $t$ of Eq.~(\ref{timeevolution}) is given by
\begin{equation}
W(x,p,t)=\frac{1}{\pi\hbar}\int_{-\infty}^{\infty}\chi_{\beta}^{*}(x-z,t)\chi_{\beta}(x+z,t)
e^{-2ipz/\hbar}dz.
\end{equation}
The PT potential is an infinite array of identical impenetrable wells of width $\pi/2 \alpha$ and therefore the wave-function can be taken equal
to zero outside the interval  $[0,\pi/2\alpha]$. By imposing this fact, one finds that the following limits of integration can be taken in the
above integral: in the left half of the well $(0\leq x \leq\pi/4 \alpha)$, one has $-x\leq z \leq+x$, while in the right half $(\pi/4 \alpha\leq
x \leq\pi/2 \alpha)$, one has $-(\pi/2 \alpha-x)\leq z \leq(\pi/2 \alpha-x)$.

Let us now analyze the Wigner function of the time-evolved CS
wave-packet at different fractional revival times.

\subsection{Symmetric case}
First we choose the symmetric case with parameter values $\rho=\kappa=50$ and $\alpha=2\;a.u.$ The CS parameter is taken as $\beta=0.6$, which
corresponds to an initial wave-packet with an energy distribution peaked around the level $\bar{n}=12$. The Wigner functions of the CS
wave-packet at different fractional revival times in phase space are shown in Fig.~\ref{wigSymAll}. At time $t=T_{rev}/4$ (see
Fig.~\ref{wigSymAll}(a)) the state is a cat-like state and its Wigner function can be explicitly written in the following two equivalent ways
\begin{widetext}
\begin{eqnarray}
W(x,p,T_{rev}/4)&=&\frac{1}{2}\left[W(x,p,0)+W(\frac{\pi}{2\alpha}-x,p,0)\right]-2
Im(W_{oe}(x,p))\label{Woe}\\
&= &\frac{1}{2}\left[W(x,p,0)+W(\frac{\pi}{2\alpha}-x,p,0)\right]+\sum_{m,n}d_m d_n (-1)^m Im(W_{mn}(x,p))\nonumber .
\end{eqnarray}
\end{widetext}
The first two terms in Eq.~(\ref{Woe}) are the two localized CS Wigner functions, situated at the left and right sides of the well. The
interference fringes are associated to the third terms of the two expressions above, where the following quantities appear
\begin{eqnarray}\nonumber
W_{oe}(x,p)=\frac{1}{\pi\hbar}\int_{-\infty}^{\infty}\chi_{\beta}^{o}(x-z,0)\chi_{\beta}^{e}(x+z,0)
e^{-2ipz/\hbar}dz,\\
W_{mn}(x,p)=\frac{1}{\pi\hbar}\int_{-\infty}^{\infty}\psi_{m}(x-z)\psi_{n}(x+z,0)
e^{-2ipz/\hbar}dz.\nonumber
\end{eqnarray}
The oscillatory interference ripples are parallel to the line joining the two split sub-CSs, i.e., the interference pattern is oscillatory only
in $p$ space and also squeezed. The number of ripples decreases as the two sub-CSs come closer.

At $T_{rev}/8$ (Fig.~\ref{wigSymAll}(b)), the initial wave-packet
splits into four distinct sub-wave-packets
(Eq.~(\ref{oneeightcl})). The Wigner function contains four
distinct parts $W_{1},W_{2},W_{3}$ and $W_{4}$ (see
Fig.~\ref{wigSymAll}(b)) and six interference terms, among which
two diagonal partners ($W_{13}$ and $W_{24}$) overlap at the
center of the phase space and generate a smaller chess-board
interference pattern. This structure is the signature of
sub-Planck scale structures \cite{zurek,ghosh} for a compass-like
state. The central interference pattern can also be seen as the
superposition of interferences of two orthogonally situated cat
states, and it is formed by small ``tiles'' much smaller than the
individual CS peaks.

\begin{figure*}[htbp] \centering
  \includegraphics[width=7.in]{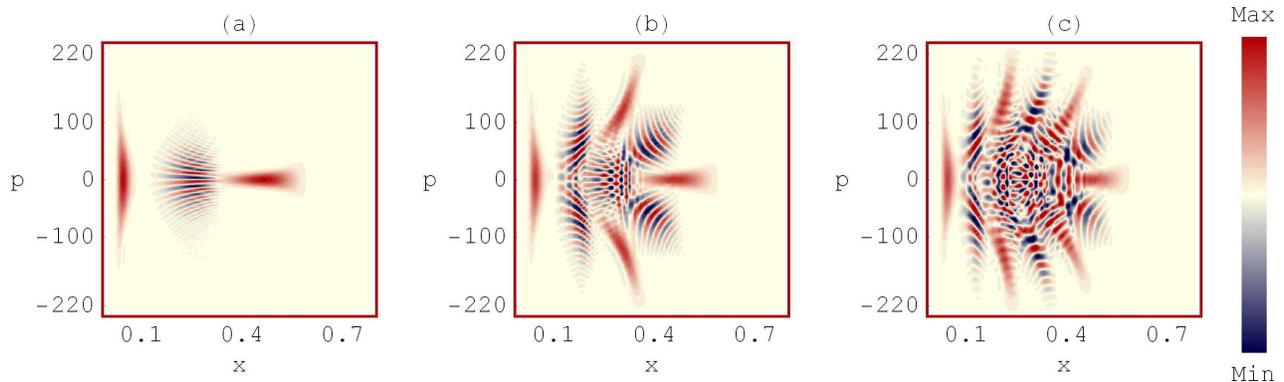}
\caption{(Color online) Contour plots of the Wigner distribution
of the time evolution of the CS, at different fractional revival
times, for a strongly asymmetric PT potential. (a) cat-like state
at $T_{rev}/4$; (b) compass-like state at $T_{rev}/8$; (c)
benzene-like state at $T_{rev}/12$. We have chosen $\rho=50$,
$\kappa=6$, $\alpha=2\;a.u.$, $\beta=0.88$, and $x$, $p$ are the
conjugate variables in atomic units.} \label{wigAsym}
\end{figure*}

Finally, at $t=T_{rev}/12$, the initial wave-packet splits into six CSs (i.e., three pairs of cat-like states), forming a benzene-like structure
(Fig.~\ref{wigSymAll}(c)). Sub-Planck structures also appear here in the center due to the superposition of three interference regions. A visual
analysis shows that the dimension of the sub-Planck scale ``tiles'' of the compass-like state (Fig.~\ref{wigSymAll}(b)) is smaller than that of
the benzene-like state (Fig.~\ref{wigSymAll}(c)). The reason behind it is that the area of overlap increases with the number of split CSs, but
we will perform a more quantitative study in the following subsection.

\subsection{Asymmetric case}

We now discuss sub-Planck scale structures for a strongly
asymmetric potential. The potential parameter values are taken as
$\rho=50$, $\kappa=6$ and $\alpha=2\;a.u.$, as in
Fig.~\ref{potential} (dashed line). For the sake of comparison
with the symmetric case, we choose the CS parameter
($\beta=0.88$), such that the energy distribution is again peaked
around $\bar{n}=12$, as in the symmetric case. In spite of this,
the interference structures are now very different. The features
of the case with large asymmetry is illustrated in
Fig.~\ref{wigAsym}. The cat-like, compass-like, and the
benzene-like states become asymmetric and inclined towards the
left boundary of the well. Since the potential well is of infinite
depth, the stiffness of the potential stretches or squeezes the
wave-packet near the left side of the well. Moreover it spreads in
$x$ space and becomes more squeezed in $p$ space. The asymmetry
affects also the interference pattern of the states: the
interference ripples are not anymore placed at the center, the
diagonal interference patterns overlap partially and destroy the
chess-board and benzene-like structures. Moreover, the
interference pattern and the localized sub-CSs partially overlap
and become much less distinguishable.

\subsection{Quantitative analysis of the sub-Planck scale structures}

Let us now make a quantitative analysis and check if the smallest
structures in the interference patterns, e.g., the alternate
positive and negative tiles have really sub-Planck scale
dimensions. Therefore we have to verify if the phase-space area
occupied by the tiles `$a$' scales, as predicted, as
$\sim\hbar^2/A$ ($\sim 1/A$ in atomic units), where $A$ is the
classical action of the state. $A$ is approximately given by the
product of the effective support of its state in position and
momentum: $A\sim \Delta x\times \Delta p$, where $\Delta
x=\sqrt{\langle x^2 \rangle - \langle x \rangle^2}$ and $\Delta
p=\sqrt{\langle p^2 \rangle - \langle p \rangle^2}$.

\begin{table*}[htbp]\centering
\caption{First row: Classical action $A$, calculated from the uncertainty product; second row: size of the phase-space sub-Planck scale
structure ($a$ in $a.u.$) in the compass-like state ($T_{rev}/8$) for the symmetric potential ($\rho=50$,$\kappa=50$). $\beta$ varies from $0.3$
to $0.8$.} \vskip .1in
\begin{tabular}{|c||c|c|c|c|c|c|c|c|c|c|c|}
  \hline
   \scriptsize{Classical Action,}& 0.748 & 0.867 & 1.189 & 1.904 & 3.175 & 5.58 & 9.225 & 13.123 & 18.349 & 23.256 & 30.03\\
   [-1.0ex]
   \scriptsize{$A\sim\Delta x \Delta p$} &  &  &  &  &  &  &  &  &  &  &  \\
  \hline
  \scriptsize{Area of a single tile ($a$),} & 1.41 & 1.2 & 0.859 & 0.57 & 0.332 & 0.194 & 0.11 & 0.078 & 0.057 & 0.045 & 0.035 \\
  [-1.0ex]
  \scriptsize{from Wigner plot} &  &  &  &  &  &  &  &  &  &  &  \\
  \hline
\end{tabular}
\label{scaling}
\end{table*}

\begin{figure}[htbp] \centering
\includegraphics[width=3.in]{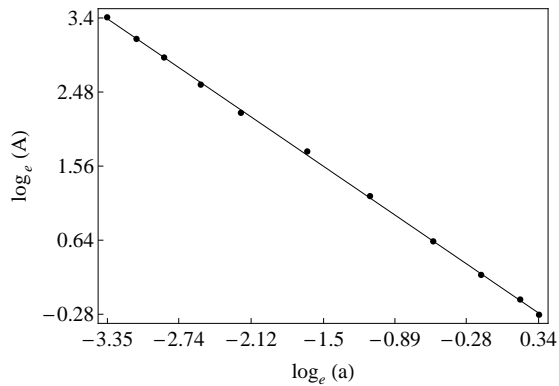}
\caption{$Log_e$-$log_e$ plot of the classical action $A$ ($a.u.$)
against the area of sub-Planck scale structures ($a$ in $a.u.$) in
the Wigner function of the compass-like state for the symmetric PT
potential with $\rho=\kappa=50$. The expected scaling law, $a\sim
1/A$, is well verified, since the data are well fitted by the line
with slope $-1.02$.} \label{loglog}
\end{figure}

In order to verify the scaling law, we first choose a symmetric
potential ($\rho=\kappa=50$) and study the interference structures
for CSs with different weighting distributions. More precisely, we
vary the parameter $\beta$ from $0.8$ to $0.3$, which corresponds
to change the position of the peak of the energy distribution from
$n=32$ to $n=2$. The classical action progressively decreases for
decreasing CS mean energy, while at the same time the size of the
sub-Planck interference structure becomes larger. This behavior is
quantitatively confirmed by Table~I, where the first row gives the
total phase space area measured form the uncertainty product,
while the second row shows the area of the structures measured
from the Wigner plots. These data are plotted in the
$log_e$-$log_e$ plot of Fig.~\ref{loglog}, where they are well
fitted by a straight line of slope $-1.02$. This latter plot
clearly confirms the expected scaling $a\sim 1/A$ with a very good
approximation, confirming also that the interference pattern has
sub-Planck dimensions.

We have then extended the analysis of the size of the sub-Planck structures to situations with different symmetry, by comparing the expected
dimension of the tiles from the measured phase-space area $A$ ($a \sim 1/A$), and the actual size of the tiles directly obtained from the plot
of the Wigner function, for the case of the asymmetric PT potential, and for the benzene state evolved up to time $T_{rev}/12$. The results are
illustrated in Table~II. We have compared the compass-like and the benzene-like states for the symmetric PT potential (rows $1-2$) and for two
different asymmetric cases (rows $3-6$). We find a good agreement between the expected and the actual dimension of the sub-Planck tiles in all
six cases and this fact suggests that the scaling $a \sim \hbar^2/A$ remains valid also in the asymmetric potential and for the benzene-like
state, even though the range of values of $a$ considered in Table~II is too small to allow us to make a firm quantitative statement \cite{note}.
We find the smallest phase-space sub-Planck structure for the compass-like state of the symmetric potential (row 1 and Fig.~\ref{wigSymAll}(b)).
This is a consequence of two facts that can be easily inferred from our analysis: i) the phase-space area of the tiles is always larger in the
case of the benzene-like state; ii) the phase-space area of the tiles \emph{increases} for increasing asymmetry of the PT potential. In the next
Section, following the results of \cite{toscano,dalvit} we will study the sensitivity to perturbations of the PT sub-Planck structures. One
expects to have the highest sensitivity with a state possessing the finest structures in phase space and therefore the above results suggest to
choose the compass-like state in the symmetric potential case for this study.

\begin{figure*}[htbp] \centering
\includegraphics[width=5.2in]{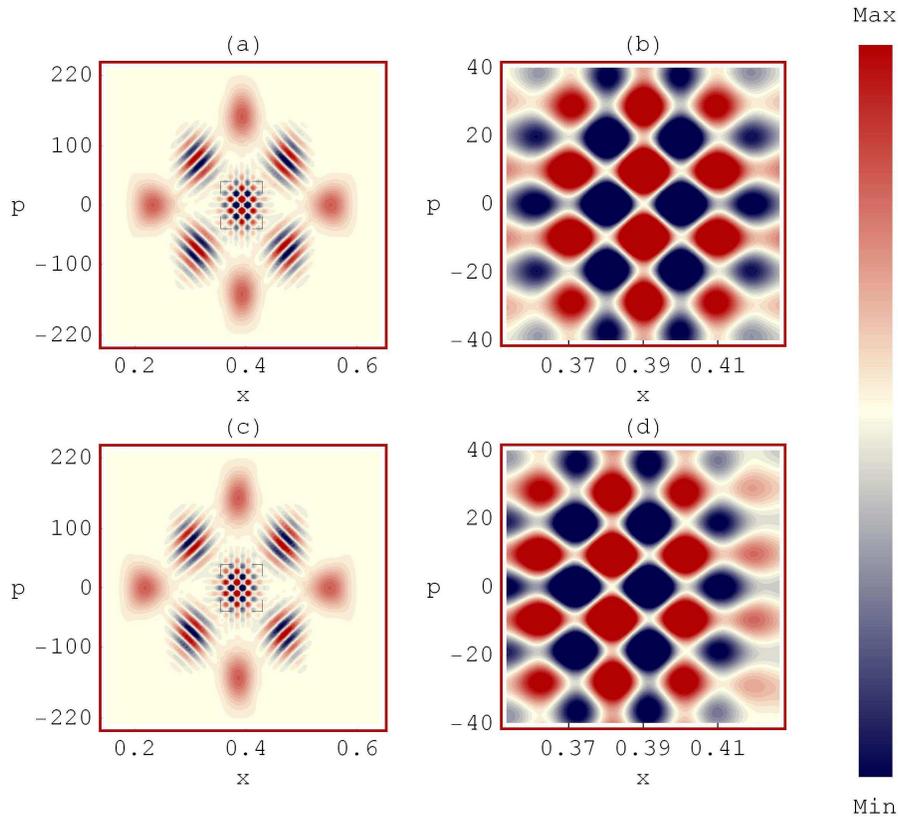}
\caption{(Color online) Contour-plot of the Wigner function of the
compass-like state for the (a) symmetric ($\rho=\kappa=50$) and
(c) slightly asymmetric ($\rho=50$, $\kappa=46$) potentials.
Enlarged views of the central interference regions of (a) and (c)
are shown in (b) and (d) respectively. Here $\alpha=2\;a.u.$ and
$x$, $p$ are the conjugate variables in atomic units.}
\label{planck}
\end{figure*}
\vskip -.02in
\begin{table}[htbp]\centering
\caption{Classical action $A$ and estimated dimension of the
sub-Planck scale structure for the compass-like ($T_{rev}/8$) and
benzene-like state ($T_{rev}/12$), both in the symmetric (rows
1-2) and in the asymmetric potential case (rows 3-6).}
\vskip .1in
\begin{tabular} {|c|c|c|c|c|c|}
\hline
  &  &   & \scriptsize{Time} &  \scriptsize{Dimension of sub-}  & \scriptsize{Area of a single }   \\
  [-1.0ex]
  &   $\rho$  & $\kappa$  & \scriptsize{(scaled by} &  \scriptsize{Planck structures}& \scriptsize{tile, from Wigner}    \\
  [-1.0ex]
  &   &   &  $T_{r\!e\!v}$)& \scriptsize{($a \sim \! 1/A$, in $a.u.$)}& \scriptsize{plot (in $a.u.$)}   \\
 \hline \hline
1 &\;\; 50 \;\; & \;\; 50 \;\; & \;\; 1/8 \;\; & \;\; 0.1084    \;\; & 0.110    \\
2 & 50 & 50 & 1/12 & 0.1420 &  0.144  \\
\hline
3 & 50 & 34 & 1/8 &  0.1278 &  0.132  \\
4 & 50 & 34 & 1/12 &  0.2463 &  0.250  \\
\hline
5 & 50 & 22 & 1/8 &  0.2118 &   0.225 \\
6 & 50 & 22 & 1/12 &  0.2878 &   0.290 \\
\hline
\end{tabular}
\label{table}
\end{table}
\vskip -.07in

\section{Sensitivity of sub-Planck scale structures}

An interesting feature of sub-Planck structures is that they signal the fact that the corresponding quantum state is extremely sensitive to
perturbations. We now verify if this holds also for the PT sub-Planck structures found above. Here we consider two different kinds of
perturbation: i) an ``internal'' perturbation, corresponding to a slight change of the PT potential, which is changed from symmetric
($\rho=\kappa$) to a slightly asymmetric one ($\kappa \lesssim \rho$); ii) an ``external'' perturbation, i.e., a phase-space displacement of the
state, for example due to the application of a weak force. As discussed in the preceding section, we consider an initially symmetric PT
potential and the compass-like state at the fractional revival time $T_{rev}/8$.

\textbf{I.} We set the potential parameter values $\rho=50$,
$\kappa=46$ and $\alpha=2\;a.u.$, which give a very small
asymmetry (see the solid line in Fig.~\ref{potential}). We compare
this case with the symmetric case $\rho=\kappa=50$. We observe
that the small asymmetry does not cause an appreciable
modification on the phase space behavior of the compass-like state
as a whole (see the contour plots in Fig.~\ref{planck}(a) and
Fig.~\ref{planck}(c) for the symmetric and slight asymmetric case
respectively). However, this is no more true if we look at the
effect of the slight asymmetry on the interference pattern and we
zoom in the center of the phase space plot (see
Fig.~\ref{planck}(b) and Fig.~\ref{planck}(d)). The effect of the
potential asymmetry on the interference pattern is equivalent to a
phase-space shift, because one has a displacement of a single tile
such that, approximately, the maximum of a tile comes to coincide
with a zero of the unperturbed state. This fact just confirms that
one has an extreme sensitivity with respect to small changes of
the potential because the two states becomes quasi-orthogonal to a
very good approximation ($\sim 0.1$). In fact, one has that
\begin{equation}
|\langle\chi_{\beta}^{\mathrm{sym}}|\chi_{\beta}^{\mathrm{asym}}\rangle|^2\!\!=\!\!\int\!\!\!\!\int \!W_{\mathrm{sym}}(x,p)
W_{\mathrm{asym}}(x,p)\; dx\; dp;\label{wigoverlap}
\end{equation}
when the two interference patterns are displaced by a tile, one has a destructive interference effect, the above integration tends to zero and
the two states become quasi-orthogonal. Thus, the size of sub-Planck structures sets a sensitivity limit on probing small potential
modifications.

\begin{figure}[htpb] \centering
\includegraphics[width=3.2in]{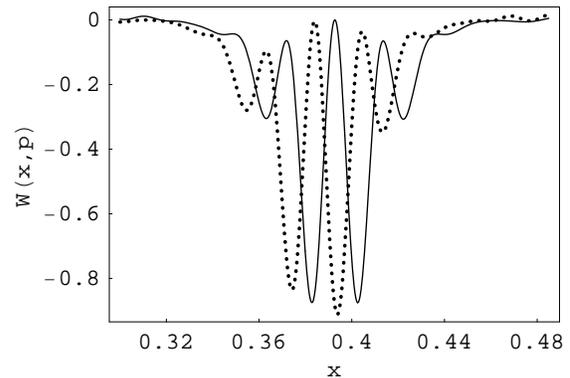}
\caption{Oscillations of the central sub-Planck region in
$x$-space at $p=0$. Solid line refers to the symmetric case
($\rho=\kappa=50$), while the dotted line refers to the slightly
asymmetric case ($\rho=50$, $\kappa=46$) as shown in
Fig.~\ref{planck}. Here, $x$ is in atomic unit.}\label{2dsmall}
\end{figure}

This sensitivity is also visualized in Fig.~\ref{2dsmall}, showing
the section at $p=0$ of the Wigner function of the states at
$T_{rev}/8$ for the two potentials. The Wigner distribution is
negative, due to the non-classicality of the state, and a shift of
the maximum of one curve to a minimum of the other is clearly
visible. This means a sub-Planck scale shift along $x$ ($\sim
1/\Delta p$) of about $0.01247\;a.u.$, consistent with that in
Fig.~\ref{planck}. A small difference in amplitudes due to the
asymmetric effect is also captured in this enlarged $2$D view.

\begin{figure}
\centering
\includegraphics[width=2.7 in]{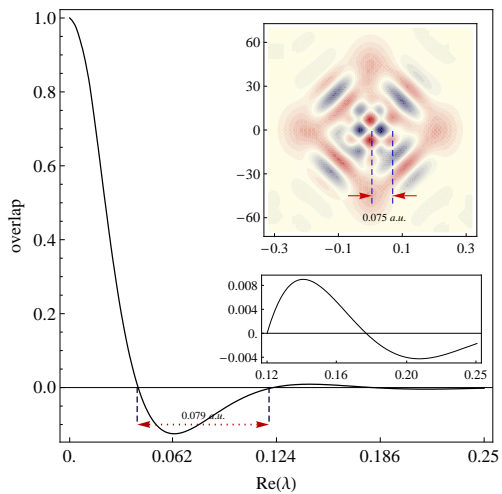}
\caption{(Color online) Plot of the overlap function of
Eq.~(\ref{overlap}) versus the real part of the displacement
parameter $\lambda$ (in $a.u.$) at the fractional revival time
$T_{rev}/8$ for $\beta=0.4$. Here $\beta^{'}=\tanh |\lambda|\;
e^{i\pi/4}$. The upper inset shows the central interference
pattern of the Wigner function and the size of the sub-Planck
structures. The lower inset shows an enlarged view of the
oscillation for larger $\lambda$. The size of the sub-Planck
structure is consistent with the period of oscillation of the
overlap.}\label{plot}
\end{figure}

\textbf{II.} The perturbation studied above is difficult to implement in practice because one has to change to potential by keeping its PT form.
In typical experimental situations a small perturbation can be applied through a weak constant force, which will physically shift the state in
phase space. Therefore we consider what happens if we shift in phase space the compass-like state of a symmetric PT potential. We apply the
displacement operator $\exp(\lambda \; K_{+} -\lambda^{*} \; K_{-})$ on the CS of Eq.~\ref{timeevolution}, where $\lambda$ is the displacement
parameter and $K_{+}$,$K_{-}$ are the SU(1,1) generators \cite{utpal}. The overlap between the initial and final states in terms of the Wigner
distribution is
\begin{equation}
|\langle\chi_{\beta}^{\mathrm{CS}}|\chi_{\beta^{'}}^{\mathrm{DCS}}\rangle|^2=\int\!\!\!\int W_{\mathrm{CS}}(x,p) W_{\mathrm{DCS}}(x,p)
dx\;dp,\label{wigoverlap1}
\end{equation}
where $\chi_{\beta}^{\mathrm{CS}}$ is the initial CS evolved at time $t$ and $\chi_{\beta^{'}}^{\mathrm{DCS}}$ is the displaced coherent state
(DCS) at the same time. After a long calculation we find the DCS and the above mentioned overlap can be written as
\begin{equation}
\langle\chi_{\beta}^{\mathrm{CS}}|\chi_{\beta^{'}}^{\mathrm{DCS}}\rangle
=\sum_{n,m,p}\textbf{D}_{m,n,p}^{\;\beta,\beta^{'}}\;\;e^{-i(E_n-E_{n-m+p})t},\label{overlap}
\end{equation}
\begin{widetext}
where
\begin{eqnarray}
\textbf{D}_{m,n,p}^{\;\beta,\beta^{'}}&=&\frac{\beta^{2n-m+p}(-\beta^{'})^{m}(\beta^{'})^{p}}{m!p!}
\times
\frac{\Gamma(2\rho+2n-m+p)\Gamma(\rho+n+\frac{1}{2})\Gamma(2\rho-m+p+1)\Gamma(\rho+n-m+p+
\frac{1}{2})}{\Gamma(n-m+1)\Gamma(\rho+n-m+\frac{1}{2})\Gamma(2\rho+p+1)\Gamma(2\rho+2n-2m+p)}\nonumber\\
&& \times
\frac{\Gamma(\rho+n-m+p+\frac{1}{2})}{\Gamma(2\rho+n-m+p)(n-m+p+\rho)}\;\;
e^{\eta(n-m+\rho/2+1/4)}.
\end{eqnarray}
\end{widetext}
Here $\lambda\!\!=\!|\lambda|e^{i\theta}$,
$\beta^{'}\!\!\!=\!\tanh|\lambda|e^{i\theta}$ and $\eta=-2\log_e
(\cosh|\lambda|)$, after using the normal form of the
disentanglement formula for SU(1,1) algebra \cite{perelomov}. We
choose a symmetric potential for $\rho=50$, $\alpha=2\;a.u.$ and
an initial CS parameter $\beta=0.4$. We have plotted the overlap
of Eq.~\ref{overlap} at time $T_{rev}/8$ (Fig.~\ref{plot}) versus
the displacement parameter $\lambda$. One has damped oscillations
which are again caused by the interference pattern of the
compass-like state. The two states becomes quickly orthogonal,
showing the sensitivity of the state to displacement. This
oscillation is similar to the one shown by the overlap between the
displaced and the undisplaced compass-like state of the harmonic
oscillator (HO) case \cite{toscano}, with the relevant difference
that here in the PT case they decay very quickly. The oscillations
for larger $\lambda$ are shown in the enlarged view in the lower
inset of Fig.~\ref{plot}. This decay of the overlap oscillations
is due to the fact that for physical systems with SU(1,1) or SU(2)
algebras as our PT system, a displaced coherent state does not
remain a coherent state, due to the higher order terms in the
Baker-Campbell-Hausdorff development of the displacement operator.
Therefore, the larger the displacement $\lambda$, the more
distorted the coherent state, implying that the overlap is always
close to zero when the displacement is not small. The HO instead
obeys the Heisenberg–Weyl algebra ($[a,a^{\dagger}]=1$) and the
displaced coherent state is still a coherent state, so that the
overlap oscillations decay much more slowly. However the important
point of our result is that, even though only one or two
oscillations are visible, the period of oscillation, which gives
the scale of quasi-orthogonality, is extremely small. This means
that the PT system is very sensitive to phase-space displacement.
Moreover this sensitivity is due just to the sub-Planck structures
of the compass-state. In fact, the period of the overlap
oscillation is $0.079\;a.u.$ (see the dotted arrow in
Fig.~\ref{plot}), which is very close to the $x$-span of a
sub-Planck structure of the compass-like state, equal to
$0.075\;a.u.$, as shown by zoomed view of the Wigner function in
the upper inset of Fig.~\ref{plot}. This means that shifting a
single sub-Planck tile results in the cancellation between the
positive and negative contributions of the two Wigner functions,
thereby making the states quasi-orthogonal.

\section{Conclusions}
We have shown that sub-Planck scale structures emerge in the time
evolution at fractional revival times also in the PT potential. We
have studied in particular the generation of compass-like states
at $T_{rev}/8$ and benzene-like states at $T_{rev}/12$. We have
studied both the symmetric and strongly asymmetric PT potential
and we have verified that in all cases the phase-space area of the
sub-Planck structures scales as $\hbar^2/A$, where $A$ is the
classical action associated with the state. In particular we have
seen the smallest phase-space structure (`tile') is associated
with that of the compass-like state of the symmetric PT potential.
For this reason we have considered this state for analyzing the
sensitivity of the PT system to perturbations. We have considered
the effect on this state of a slight asymmetry of the potential
and of a phase-space displacement. We have seen that in both
cases, the sub-Planck structures are responsible for high
sensitivity, because as soon as a single tile is displaced in
phase space so that its maximum coincides with a minimum of the
undisplaced state, one has destructive interference and the two
states become approximately orthogonal.


\begin{thebibliography}{99}
\bibitem{note}The numerical study of a wider interval of values for `$a$' turns out to be problematic in the case of asymmetric PT potentials
because for larger asymmetry the direct measurement of `$a$' from
the Wigner plots became more and more unreliable.
\end{thebibliography}
\end{document}